\documentclass[12pt,aps,prb,floats,eqsecnum,graphicx,euscript,amsmath] {revtex4}
\usepackage[dvips]{graphicx}
\begin{document}
\bibliographystyle{prsty}
\title{
{Zero-Momentum Cyclotron Spin-Flip Mode in a Spin-Unpolarized
Quantum Hall System}} \vskip -0.mm
\author{S.~Dickmann, and I.V.~Kukushkin}

\affiliation{Institute for Solid State Physics, Russian Academy of
Scieces, Chernogolovka, 142432 Russia }

\begin{abstract}
We report on a study of the zero-momentum cyclotron spin-flip excitation
in the ${\mbox{\scriptsize ${\cal V}$}}\!=\!2$ quantum Hall regime.
Using the excitonic representation the excitation energy is calculated
up to the second order Coulomb corrections. A considerable negative
exchange shift relative to the cyclotron gap is established for
cyclotron spin-flip excitations in the the spin-unpolarized electronic
system. Under these conditions this type of states presents the {\it
lowest-energy} excitations. For a fixed filling factor
(${\mbox{\scriptsize ${\cal V}$}}\!=\!2$) the energy shift is
independent of the magnetic field which is in agreement with recent
experimental observations. \vspace{1.mm}


\end{abstract}

\maketitle

\def\bbf{\bf}

\bibliographystyle{prsty}

\vskip 7mm

{\bf 1}. It is well known that in a translationally invariant
two-dimensional electron system Kohn's theorem$\,$ \cite{ko61} prohibits
coupling of a homogeneous external perturbation to collective
excitations of the electrons. As a result, the energy of cyclotron
excitations (CE) at zero-momentum has no contribution from Coulomb
interaction and the dispersion of CE starts from the cyclotron gap. In
addition to inter-Landau-levels cyclotron excitations [magnetoplasma
(MP) mode] there are two other branches of collective excitations in the
system of 2D-electrons: intra-Landau levels spin-flip (SF) excitations
(spin-waves) and inter-Landau-levels combined cyclotron spin-flip
excitations (CSFE's). In the case of SF excitations, there exists
Larmor's theorem which forbids any contribution from Coulomb interaction
to the excitation energy at zero-momentum in spin rotationally invariant
systems (see, e.g., Ref. \onlinecite{ka84}). However, in contrast to the
CE and SF excitations, there are no symmetry reasons for the absence of
many-body corrections to the zero-momentum energy of CSFE's. Moreover,
it is well established now both theoretically and experimentally$\,$
\cite{pi92} that for the spin-polarized electron system
(${\mbox{\footnotesize ${\cal V}$}}\!=\!1$) the energy of cyclotron
spin-flip excitations is strongly shifted to higher values relative to
the cyclotron gap due to the exchange interaction. Therefore, the energy
of combined cyclotron spin-flip excitations is a very convenient tool to
probe many-body effects, for example, in the inelastic light scattering
measurements performed at zero momentum. The sensitivity of CSFE energy
at ${\bf q}=0$ to many-body effects strongly depends on the spin
polarization of the electron system. For the spin-unpolarized electron
system (${\mbox{\footnotesize ${\cal V}$}}\!=\!2$),
theory$\,$\cite{ka84} developed within the first order perturbation
approach in terms of the parameter $r_c=E_C/\hbar\omega_c$ ($E_C$ is the
characteristic Coulomb energy, $\omega_c$ is the cyclotron frequency)
predicts a zero many-body contribution to the zero-momentum energy of
CSFE. This result is in contradiction with recent experimental data.
\cite{kul04}  We show below that calculation of the CSFE zero-momentum
energy of for the ${\mbox{\footnotesize ${\cal V}$}}\!=\!2$ system
performed to within the second order Coulomb corrections yields a
considerable negative exchange shift relative to the cyclotron gap.

The studied system is characterized by exact quantum numbers $S$, $S_z$
and ${\bf q}$ and by a non-exact but `good' quantum number $\delta n$
corresponding to the change of the single-electron energy
$\hbar\omega_c\delta n$ with an excitation. The relevant excitations
with ${\bf q}\!=\!0$ and $\delta n\!=\!1$ may be presented in the form
${\hat{K}}_{S,S_z}^{\dag}|{\bf 0}\rangle$, where $|{\bf 0}\rangle$ is
the ground state and ${\hat{K}}_{S,S_z}^{\dag}$ are ``raising"
operators:
$\!{\hat{K}}_{0,0}^{\dag}\!\!=\!\!\sum_{np\sigma}\!{}\!\sqrt{n+1}
c^{\dag}_{n\!+\!1,p,\sigma}c_{n,p,\sigma},
\;{\hat{K}}_{1,0}^{\dag}\!\!=\!\!\sum_{np\sigma}\!{}\!\sqrt{n+1}\,
(-1)^{\sigma}c^{\dag}_{n\!+\!1,p,\sigma}c_{n,p,\sigma}
\qquad\mbox{and}\qquad{\hat{K}}_{1,+/-}^{\dag}=\sum_{np}\!{}\!
\sqrt{n+1}c^{\dag}_{n\!+\!1,p,\uparrow/\downarrow}
c_{n,p,\downarrow/\uparrow}$, [$c_{n,p,\sigma}$ is the Fermi
annihilation operator corresponding to the Landau-gauge state $(n,p)$
and spin index $\sigma\!\!=\!\uparrow,\!\downarrow$]. The commutators
with the kinetic-energy operator ${\hat{H}}_1$  are
$[{\hat{H}}_1,{\hat{K}}_{S,S_z}^{\dag}]\!\!\equiv
\!\!\hbar\omega_c{\hat{K}}_{S,S_z}^{\dag}$. (The total Hamiltonian is $
{\hat H}_{\mbox{\scriptsize tot}}\!=\!{\hat H}_1\!+\!{\hat
H}_{\mbox{\scriptsize int}}$, where ${\hat H}_{\mbox{\scriptsize int}}$
is the exact Coulomb-interaction Hamiltonian.) If $|{\bf 0}\rangle$ is
unpolarized, we have ${\hat{\bf S}}^2{\hat{K}}_{S,S_z}^{\dag}|{\bf
0}\rangle\!\!\equiv\!\!S(S\!+\!1){\hat{K}}_{S,S_z}^{\dag}|{\bf
0}\rangle$, ${\hat{S}}_z{\hat{K}}_{S,S_z}^{\dag}|{\bf
0}\rangle\!\!\equiv\!\!S_z{\hat{K}}_{S,S_z}^{\dag}|{\bf 0}\rangle$ and
besides get the identity $\langle {0}|{\hat{K}}_{S,S_z}[{\hat
H}_{\mbox{\scriptsize int}},{\hat{K}}_{S,S_z}^{\dag}]|{
0}\rangle\!\equiv\!0$ ($|0\rangle$, to describe the zero'th order ground
state). The latter determines the first-order Coulomb corrections
vanishing both for the $S\!=\!0$ MP mode and for the $S\!=\!1$ triplet
states corresponding to the combined spin-cyclotron excitation. At the
same time $[{\hat H}_{\mbox{\scriptsize
int}},{\hat{K}}_{0,0}^{\dag}]\!\equiv\!0\,$ \cite{ko61} but $[{\hat
H}_{\mbox{\scriptsize int}},{\hat{K}}_{1,S_z}^{\dag}]\!\neq\!0$ which
means that the MP mode indeed has no exchange energy calculated to {\it
any order} in $r_c$, whereas the triplet states have the exchange
correction even in terms of $r_c^2$.

The second-order correction, $\Delta E_{\mbox{\scriptsize
SF}}\!\sim\!\hbar\omega_c r_c^2$, does not depend on the magnetic
field since $E_C\!=\!\alpha e^2/\varepsilon l_B$. The
renormalization factor, $\alpha$, is determined by the
size-quantized wave function of electrons confined to the quantum
well (QW). In the ideal 2D case $\alpha\!=\!1$. However,  in
experiments with comparatively wide QW's we expect a well reduced
value of $\alpha$. Our {\it analytical} calculation of the second
order correction to the CSFE energy is performed in terms of $r_c$
assumed to be small.

All three triplet states have certainly the same exchange energy,
and it is sufficient to calculate this, e.g., for the CSFE with
$S\!=\!1$ and $S_z\!=\!-\!1$. The obtained result confirms
experimental observations.

{\bf 2}. The most adequate approach to the integer-quantum-Hall
calculations is based on the Excitonic Representation (ER)
\cite{di02,dz83} technique (see also, e.g., Refs.
\onlinecite{di96}). The latter means that instead of
single-electron states belonging to a continuously degenerate
Landau level (LL) we employ the exciton states ${\cal Q}_{ab\,{\bf
q}}^{\dag}|0\rangle$ as the basis set. Here $|0\rangle$ is the
ground state found in the zero approximation in $r_c$ (it remains
also the same even calculated within the framework of the mean
field approach). The exciton creation operator is defined as$\,$
\cite{dz83,di02,di96,by87}
$$\vspace{-2.mm}
  {\cal Q}_{ab\,{\bf q}}^{\dag}=\frac{1}{\sqrt{ N_{\phi}}}\sum_{p}\,
  e^{-iq_x p}
  b_{p+\frac{q_y}{2}}^{\dag}\,a_{p-\frac{q_y}{2}}\,. \eqno
  (1)\vspace{-12.mm}
$$
$N_{\phi}=A/2\pi l_B^2$ stands for the number of magnetic flux
quanta, ${\bf q}\!=\!(q_x,q_y)$ is the 2D wave vector in units of
$1/l_B$. Binary indexes $a$ and $b$ present both the LL number and
spin index. [I.e. $a\!=\!(n_a,\sigma_a)$, and $a_p$ in Eq. (1)
stands for the corresponding annihilation operator; when
exploiting below the notation $a\!=\!n$ or $a\!=\!{\overline n}$
as sublevel indexes, this means that $a\!=\!(n,\uparrow)$ or
$a\!=\!(n,\downarrow)$, respectively.] The annihilation exciton
operator is ${\cal Q}_{ab\,{\bf q}}\!\equiv\!{\cal Q}_{ba\,-\!{\bf
q}}^{\dag}$. The commutation rules define a special Lie algebra:
\cite{di02,di96,by87}
$$
\begin{array}{l}
{}\!\!{}\!{}\!{}\!{}\!{}\!{}\!\left[{\cal Q}_{cd\,{\bf
q}_1}^{\dag},{\cal Q}_{ab\,{\bf
   q}_2}^{\dag}\right]\!\equiv\! N_{\phi}^{-1/2}\left[  e^{-i({\bf
   q}_1\!\times\!{\bf q}_2)_z/2}\delta_{b,c}{\cal Q}_{ad\,{\bf q}_1\!+\!{\bf
   q}_2}^{\dag}\right.\\
   \qquad{}\qquad{}\quad{}-\!\left.e^{i({\bf
   q}_1\!\times\!{\bf q}_2)_z/2}\delta_{a,d}
   {\cal Q}_{cb\,{\bf q}_1\!+\!{\bf
   q}_2}^{\dag}\right],
   \end{array} \eqno (2)
$$
where
$\delta_{a,b}\!=\!\delta_{n_a,n_b}\delta_{\sigma_a,\sigma_b}$ is
the Kronecker symbol. In the ${\mbox{\footnotesize ${\cal
V}$}}\!=\!2$ case we get the following identity: $
N_{\phi}^{-1/2}{\cal Q}_{aa\,{\bf
q}}^{\dag}|0\rangle\!\equiv\!\delta_{{\bf
q},0}\left(\delta_{a,0}\!+\!\delta_{a,{\overline
0}}\right)|0\rangle$.

The advantage of the exciton states lies in the fact that an
essential part of the Coulomb interaction Hamiltonian may be
diagonalized in this basis. In the perturbative approach the
excitonically diagonalized part ${\hat H}_{\mbox{\scriptsize ED}}$
should be included into the unperturbed Hamiltonian ${\hat
H}_0={\hat H}_1\!+\!{\hat H}_{\mbox{\scriptsize ED}}$ and only the
off-diagonal part ${\hat {\cal H}}_{\mbox{\scriptsize int}}\!=\!
{\hat H}_{\mbox{\scriptsize int}}\!-\!{\hat H}_{\mbox{\scriptsize
ED}}$ is considered as a perturbation. \cite{di02} In the
excitonic basis the LL degeneracy becomes well lifted because now
there are Coulomb corrections (depending on the ${\bf q}$ modulus)
to the energies of the basis states. It is useful to take into
account that all terms of the  relevant ${\hat {\cal
H}}_{\mbox{\scriptsize int}}$ part may be presented in the form
(cf.  Ref. \onlinecite{di02}) \vspace{-1.mm}
$$
  \begin{array}{l}
  {}\!{}\!{}\!{}\!{\hat{\cal  H}}_{\mbox{\scriptsize {int}}}=
  {\displaystyle \frac{e^2}{2\varepsilon l_B}}\sum_{{\bf q},a,b,c,d}
  V({\bf q})\left[h_{n_an_b}({\bf q})\delta_{\sigma_a,\sigma_b}
  {{\cal Q}}_{ab{\bf
  q}}^{\dag}\right]\\{}\quad{}\qquad{}\qquad{}\qquad{}
  \qquad\times\!\left[h_{n_cn_d}(-{\bf q}) \delta_{\sigma_c,\sigma_d}
  {{\cal Q}}_{cd\,-\!{\bf q}}^{\dag}\right].
  %\quad (3)
  \end{array}  \vspace{-1mm}\eqno (3)
$$
Here $ 2\pi V({\bf q})$ is the dimensionless 2D Fourier component
of the averaged Coulomb potential (in the ideal 2D case
$V\!=\!1/q$), and $h_{kn}({\bf q})=({k!}/{n!})^{1/2}e^{-q^2/4}
  (q_-)^{n\!-\!k}L^{n\!-\!k}_{k}(q^2/2)$ are the ER ``building-block''
functions ($L_k^n$ is the Laguerre polynomial,
$q_{\pm}=\mp\frac{i}{\sqrt{2}}(q_x\pm iq_y)$;
 cf. also Ref. \onlinecite{ka84} and Refs.  \onlinecite{di02,di96}). The
functions $h_{kn}$ satisfy the identity: $h_{kn}^*({\bf
q})\!\equiv\!h_{nk}(-{\bf q})$.

At the ${\mbox{\footnotesize ${\cal V}$}}\!=\!2$ filling the CSFE
state calculated within the zero order in ${\hat {\cal
H}}_{\mbox{\scriptsize int}}$ is simply
$|SF\rangle{}\!=\!{\hat { Q}}_{{0}\overline{1}}^{\dag}|0\rangle$
(the notation $ {\hat {\cal Q}}_{ab\,{\bf 0}}^{\dag}\!=\!
{\hat{Q}}_{ab}^{\dag}$ is employed).  We thus have ${\hat
H}_{\mbox{\scriptsize
ED}}Q^{\dag}_{{0}\overline{1}}|0\rangle\!=\!\langle 0|
 {\cal Q}_{ab\,{\bf q}}{\hat {\cal  H}}_{\mbox{\scriptsize
int}}Q^{\dag}_{{0}\overline{1}}|0\rangle\!=\!0$ for any indexes
${\bf q}$ and $ab$ (one could check it directly using the ER
approach; see also Ref. \onlinecite{ka84}). Action of ${\hat {\cal
H}}_{\mbox{\scriptsize int}}$ on the $|SF\rangle$ state leads to
two- and or even three-exciton states. Therefore the excitonic
basis should be extended.

{\bf 3}. In principle, there are eight different kinds of possible
two-exciton states at ${\mbox{\footnotesize ${\cal V}$}}\!=\!2$.
In our case the relevant ones are those corresponding to spin
numbers $S_z\!=\!-1$ and $S\!=\!0$, namely: $|\nu,1\rangle\!=\!
{\cal Q}_{0n_2\,-\!{\bf q}_{\nu}}^{\dag}\! {\cal
Q}_{0\overline{n}_1\,{\bf q}_{\nu}}^{\dag}|0\rangle$,
$|\nu,2\rangle\!=\! {\cal Q}_{\overline{0}\overline{n}_2\,-\!{\bf
q}_{\nu}}^{\dag}\! {\cal Q}_{0\overline{n}_1\,{\bf
q}_{\nu}}^{\dag}|0\rangle$, $|\nu,3\rangle\!=\! \frac{1}{2}{\cal
Q}_{0n_2\,-\!{\bf q}_{\nu}}^{\dag}\! {\cal Q}_{0n_1\,{\bf
q}_{\nu}}^{\dag}|0\rangle$, $|\nu,4\rangle\!=\! {\cal
Q}_{0n_2\,-\!{\bf q}_{\nu}}^{\dag}\! {\cal
Q}_{\overline{0}\overline{n}_1\,{\bf q}_{\nu}}^{\dag}|0\rangle$,
and $|\nu,5\rangle\!=\!\frac{1}{2} {\cal
Q}_{\overline{0}\overline{n}_2\,-\!{\bf q}_{\nu}}^{\dag}\! {\cal
Q}_{\overline{0}\overline{n}_1\,{\bf q}_{\nu}}^{\dag}|0\rangle$
(certainly only the states with zero total momentum should be
considered). We have used here $\nu$ as a composite index
corresponding to the set $(n_1,n_2, {\bf q}_{\nu})$. The
two-exciton states of different types are orthogonal, i.e.
$\langle I,\nu|\mu,J\rangle\!=\!0$ if $I\!\neq\!J$ [$\mu$ is the
set $(m_1,m_2,{\bf q}_{\mu})$, below $\lambda=(l_1,l_2,{\bf
q}_{\lambda})$,...]. However, within the same type their
orthogonalization rules should be defined in a special way.

First, let us consider a combination \vspace{-2.mm}
$$
  \sum_{\nu}f_{\nu}|\nu,I\rangle\vspace{-3.mm}
\eqno (4)
$$
(summation is performed over all components of the composite
index). In this case the function $f_{\nu}=f(n_1,n_2,{\bf
q}_{\nu})$ formally turns out to be non-uniquely defined because
only a certain transform of this has a physical meaning. Indeed,
actually only a projection of the sum (4) onto a certain
two-exciton state $|\mu,J\rangle$ would be of any sense. With the
help of commutation rules we obtain
$$
  \sum_{\nu}f_{\nu}\langle J,\mu|\nu,I\rangle\equiv\delta_{I,J}\{f_{\mu}\}_I\,,
  \vspace{-2mm} \eqno (5)
$$
(cf. Ref. \onlinecite{di02}). Here the curly brackets mean the
transform
$$
\begin{array}{l}
 {}\!{}\!{}\!{}\!{}\!{}\!{}\!{}\!{}\!\left\{f_{\nu}\right\}_I\!=\!f_{\nu}\!-
  \!{N_{\phi}}^{-1}\!\sum_{\lambda}{\cal
  F}^{(I)}_{\nu\lambda}f_{\lambda},\quad \mbox{if}\quad I=1,\,2\;\;
\mbox{or}\;4;\\
{}\!{}\!{}\!{}\!{}\!{}\!{}\!{}\mbox{and}\;\left\{f_{\nu}\right\}_{I}\!=\!\frac{1}{2}\left(f_{\nu}\!-
\!{N_{\phi}}^{-1}\!\sum_{\lambda}{\cal
F}^{(I)}_{\nu\lambda}f_{\lambda}\right),\; \mbox{if}\; I=3\;
\mbox{or}\; 5.
\end{array}  \eqno (6)
$$
The definition of the kernels ${\cal
  F}^{(I)}_{\nu\lambda}f_{\lambda}$ is also parametrized by the
  kind $I$ of the state, namely:
$$
\begin{array}{l}
{\cal F}^{(1)}_{\nu\lambda}\!=\!{\cal F}^{(3)}_{\nu\lambda}\!=
\!{\cal F}^{(5)}_{\nu\lambda}\!\equiv\!
\delta_{n_1,l_1}\delta_{n_2,l_2}e^{i({\bf q}_{\nu}\!\times\!{\bf
q}_{\lambda})_z},\\ {\cal F}^{(2)}_{\nu\lambda}\!\equiv\!
\delta_{n_1,l_2}\delta_{n_2,l_1}e^{-i({\bf q}_{\nu}\!\times\!{\bf
q}_{\lambda})_z},\;\;\mbox{and}\;\;{\cal
F}^{(4)}_{\nu\lambda}\!\equiv\!0.
\end{array}\eqno (7)
$$
Note that the transform $\{...\}_I$ is to within a factor
equivalent to its double application:
$\{\{f\}_I\}_I\!=\!K_I\{f\}_I$, where $K_1\!=\!K_2\!=\!2$ and
$K_3\!=\!K_4\!=\!K_5\!=\!1$. Therefore, if we replace, e.g.,
$f_{\nu}\to
f_{\nu}\!+\!K_I\varphi_{\nu}\!-\!\{\varphi_{\lambda}\}_I$
($\varphi_{\nu}$ is an arbitrary function), then this operation
does not affect the combinations (4) and (5). So, only the
``antisymmetrized'' part $\{f_{\nu}\}_I$ contributes to the
matrix-element calculations. The origin of this feature of the
two-exciton states is related to the permutation antisymmetry of
the total wave function describing the electron system studied
(cf., e.g., Refs. \onlinecite{di02,by83}). There is also a useful
identity
$$
   \sum_{\nu}w(n_1,n_2)f^*_{\nu}\left\{g_{\nu}\right\}_{I}\!
    \equiv\!\sum_{\nu}w(n_1,n_2)\left\{f_{\nu}
   \right\}_{I}^*g_{\nu}\,,   \eqno(8)
$$
which is valid for any kinds of the transforms $\{...\}_I$ if the
function $w$ in Eq. (5) is assumed to be such that
$w(n_1,n_2)\!\equiv\!w(n_2,n_1)$. In particular, Eq. (5) gives the
equations:
$$
\langle I,\nu|\mu,J\rangle\!\equiv\!
\delta_{I\!\!,J}\{\delta_{\nu\mu}^{(I)}\}_{I},\vspace{-3.mm}\eqno
(9)
$$
where
$$
\begin{array}{l}
\delta_{\nu\mu}^{(1)}\!=\!\delta_{\nu\mu}^{(2)}\!=\!
\delta_{\nu\mu}^{(4)}\!\equiv\!\delta_{n_1,m_1}
\delta_{n_2,m_2}\delta_{{\bf q}_{\nu},{\bf q}_{\mu}}\quad
\mbox{and}\\
\delta_{\nu\mu}^{(3)}\!=\!\delta_{\nu\mu}^{(5)}\!\equiv\!
\frac{1}{2}\left(\delta_{n_1,m_1} \delta_{n_2,m_2}\delta_{{\bf
q}_{\nu},{\bf q}_{\mu}}\!\!+\!\delta_{n_1,m_2}
\delta_{n_2,m_1}\delta_{{\bf q}_{\nu},-{\bf q}_{\mu}}\right).
\end{array}
$$
Summation in the $\{\delta_{\nu\mu}^{(I)}\}_{I}$ transform is
performed over the first index: e.g.
$\{\delta_{\nu\mu}^{(1)}\}_{1}\!\equiv\!
\delta_{\nu\mu}^{(1)}\!\!-\!{\cal
F}^{(1)}_{\nu\mu}\!\!/N_{\phi}$, and so on.

{\bf 4}. The first-order corrections (in terms of ${\hat {\cal
H}}_{\mbox{\scriptsize int}}$) to the CSFE energy are presented as
an expansion over the two-exciton states $|\nu,I \rangle$ and
three-exciton states $Q^{\dag}_{{0}\overline{1}}|\nu,I\rangle$,
namely:\vspace{-1mm}
$$
\begin{array}{l}
 {}\!{}\!{}\!  |SF,\rangle\!=\!Q^{\dag}_{{0}\overline{1}}|0\rangle+\sum_{I\!=\!1,2}\sum_{\nu}
   C^{({I})}_{\nu}|\nu,I\rangle{}\quad{}\\
{}\qquad{}\qquad{}\qquad{}\qquad
    +\sum_{I\!=\!3,4,5}\sum_{\nu}
   C^{(I)}_{\nu}Q^{\dag}_{{0}\overline{1}}
   |\nu,I\rangle.
  \end{array}
   \vspace{-1mm}   \eqno(10)
$$
A regular application of the perturbative approach \cite{ll91}
leads to the following expression for the exchange correction to
the energy: $\Delta E_{\mbox{\scriptsize SF}}\!=\!\langle
SF|\hat{\cal H}_{\mbox{\scriptsize
int}}Q_{0\overline{1}}^{\dag}|0\rangle$. Substituting
$|SF,\rangle$ from Eq. (10) we see that the {\it contribution of
the two-exciton states} to the energy arises only due to the terms
of Eq. (3) which do not commute with $Q_{0\overline{1}}^{\dag}$:
\vspace{-1mm}
$$
    \Delta
    E_{1\!-\!2}\!=\!\mbox{$\sum_{I\!=\!1,2}\sum_{\nu}$}{C_{\nu}^{(I)}}^*
    \left\langle I,\nu\right|\left[\hat{\cal H}_{\mbox{\scriptsize
    int}},Q_{0\overline{1}}^{\dag}\right]
    \left|0\right\rangle\,.   \vspace{-1mm}\eqno (11)
$$
The coefficients $C_{\nu}^{(I)}$ are determined by the equations
\vspace{-1mm}
$$
   \mbox{$\sum_{\mu}$}C_{\mu}^{(I)}\langle I,\nu|\mu,I\rangle\!=\!-\langle I,\nu
   |[\hat{\cal H}_{\mbox{\scriptsize
    int}},Q_{0\overline{1}}^{\dag}]|0\rangle/\Delta_{\nu}
    \vspace{-1mm}  \eqno (12)
$$
($I\!=\!1,2$), where
$\Delta_{\nu}\!=\!\hbar\omega_c(n_1\!+\!n_2\!-\!1)\!>\!0$ stands
for the difference of the cyclotron energies in the states
$|\nu,I\rangle$ and $Q_{0\overline{1}}^{\dag}|0\rangle$.
Calculating  the commutator in Eqs. (11)-(12) [employing the rules
(2)], and then using the properties (5) and (8) of the summation
over index, we obtain
$$
     {}\!\Delta E_{1\!-\!2}\!=\!-N_{\phi}^{-1}\mbox{$\sum_{I\!=\!1,2}\;
    \sum_{\nu}$}\left\{F_{\nu}\right\}_{I}F_{\nu}^*/(n_1\!+\!n_2\!-\!1)
    \eqno (13)
$$
[in units of 2Ry${}\!=\!(e^2/\varepsilon l_B)^2/\hbar\omega_c
\!=\!m_e^*e^4/\varepsilon^2\hbar^2$], where
$$
F_{\nu}\!=\!V({\bf q}_{\nu})\left[h_{1n_1}({\bf
q}_{\nu})\!-\!\delta_{1,n_1}h_{00}({\bf
q}_{\nu})\right]h_{0n_2}(-{\bf q}_{\nu}).\eqno (14)
$$

Now we calculate the contribution $\Delta E_{3\!-\!5}$  which is
determined by the {\it three-exciton states} [see Eq. (10)]. This
correction arises from the commuting part (with
$Q_{0\overline{1}}^{\dag}$) of $\hat{\cal H}_{\mbox{\scriptsize
int}}$ acting on the state
$Q_{0\overline{1}}^{\dag}\left|0\right\rangle$, i.e.
$$
    \Delta E_{3\!-\!5}\!=\!\sum_{I\!=\!3,4,5}\sum_{\nu}{C_{\nu}^{(I)}}^*
    \left\langle I,\nu\right|Q_{0\overline{1}}Q_{0\overline{1}}^{\dag}
    \hat{\cal H}_{\mbox{\scriptsize
    int}}\left|0\right\rangle\,.
\vspace{-1mm}\eqno (15)
 $$
The equations for the coefficients are
$$
\begin{array}{l}
 {}\!{}\!   \sum_{\mu}C_{\mu}^{(I)}\left\langle I,\nu|Q_{0\overline{1}}
   Q_{0\overline{1}}^{\dag}|\mu,I\right\rangle\qquad{}\qquad{}\qquad{}\\
 \qquad{}\qquad{}\qquad{} =\!-\left\langle I,\nu
    |Q_{0\overline{1}}Q_{0\overline{1}}^{\dag}\hat{\cal H}_{\mbox{\scriptsize
     int}}|0\right\rangle/\tilde{\Delta}_{\nu}
\end{array}  \vspace{-2mm}    \eqno (16)
$$
({I}=3,4,5), where
$\tilde{\Delta}_{\nu}\!=\!\hbar\omega_c(n_1\!+\!n_2)\!\geq 2$.
Substituting $Q_{0\overline{1}}Q_{0\overline{1}}^{\dag}\!\equiv\!
N_{\phi}^{-1/2}\left(Q_{00}\!-\!Q_{\overline{1}\overline{1}}\right)\!+\!Q_{0\overline{1}}^{\dag}
Q_{0\overline{1}}$ into Eqs. (15)-(16) we deduce that the operator
$Q_{0\overline{1}}^{\dag}Q_{0\overline{1}}$ gives no contribution,
whereas action of the remaining terms reduces the convolutions in
Eqs. (15)-(16) to the ``bra-ket'' products of two-exciton states.
In so doing we find a huge contribution (eventually $\sim
N_{\phi}$) into Eq. (15) due to the commuting part of
$N_{\phi}^{-1/2}\!\!\left(Q_{00}\!-\!Q_{\overline{1}\overline{1}}
\right)$, which is actually nothing else but the second order
correction (in terms of $r_c$) to the ground state, namely:
$\Delta E_0\!=\!\sum\limits_{\nu;\,I=3,4,5} C_{\nu}^{(I)} \langle
0|\hat{\cal H}_{\mbox{\scriptsize int}}|\nu,I\rangle$. According
to Eq. (16)\vspace{-1.mm}
$$
    \left\{C_{\nu}^{(I)}\right\}_I\!=\!-\!
    [e^2/(\varepsilon l_B\hbar\omega_c)]
    \left\{G_{\nu}\right\}_I/(n_1\!+\!n_2)
    \eqno (17)\vspace{-1.mm}
$$
(I=\!3,4,5) with\vspace{-2.mm}
$$
{}\!G_{\nu}\!\!=\!\!V\!({\bf q}_{\nu})h_{0n_1}\!({\bf
q}_{\nu})h_{0n_2}\!(-{\bf q}_{\nu})\,.  \eqno (18)\vspace{-1.mm}
$$
The non-commuting part determines the corrections to the
bra-vectors in Eq. (15). For the $I\!=\!3$ states  we get
$$
N_{\phi}^{-1/2}\!\left[Q_{00}\!-\!Q_{\overline{1}\overline{1}},\frac{1}{2}
{\cal Q}_{0n_2\,-{\bf q}_{\nu}}^{\dag}\!{\cal Q}_{0n_1\,{\bf
q}_{\nu}}^{\dag}\right]|0\rangle\!=\!-2|\nu,3\rangle/N_{\phi}\,,
\vspace{-2.mm}
$$
and correspondingly
$-\!(1\!+\!\delta_{n_1,1})|\nu,4\rangle/N_{\phi}$ and
$-\!(\delta_{n_1,1}\!+\!\delta_{n_2,1})|\nu,5\rangle/N_{\phi}$ at
$I\!=\!4$ and $I\!=\!5$ [the identities (2) have been used]. The
similar corrections to the bra-vectors in Eq. (16) do not affect
the equation (17) for $C_{\nu}^{(I)}$.

The desirable exchange shift should be measured from corrected
energy of the ground-state. We keep thus in Eq. (15) only the
contribution of the non-commuting part (i.e. considering $\Delta
E_{3-5}\to \Delta E_{3-5}\!-\!\Delta E_0$). Then by substituting
Eq. (3) for $\hat{\cal H}_{\mbox{\scriptsize int}}$ into Eq. (15)
and using again the summation rules (5) and (8) we find from Eqs.
(15) and (17) the $I\!=\!3\!-\!5$ correction
$$
\begin{array}{l}
{}\!{}\!{}\!{}\Delta
E_{3\!-\!5}\!=\!\frac{1}{N_{\phi}}\sum_{\nu}\left[\left(2\!+\!
\delta_{n_1,1}\!+
    \!\delta_{n_2,1}\right)\left\{G_{\nu}\right\}_3\right.\\
{}\qquad  {}\quad{}\qquad{}\quad
+\!\left.\left(1\!+\!\delta_{n_1,1}\right)G_{\nu}\right]
    G_{\nu}^*/(n_1\!+\!n_2)
 \end{array}  \vspace{-1.5mm} \eqno (19)
$$
(in units of 2Ry). The combination with Eq. (13)
yields\vspace{-2mm}
$$
  \Delta E_{\mbox{\scriptsize SF}}=\Delta E_{1\!-\!2}\!+\!
  \Delta E_{3\!-\!5}.  \vspace{-1mm}       \eqno (20)
$$

The sum over $\nu$ in Eqs. (13) end (19) means summation over
$n_1$ and $n_2$ and the integration over ${\bf q}_{\nu}$. This is
a routine procedure and the suitable sequence of operations is as
follows. First we perform the summation over all of $n_{1}\ge 1$
and $n_{2}\ge 1$ keeping the sum $n_{\nu}\!=\!n_{1}\!+\!n_{2}$
fixed. Then we  make the integration over ${\bf q}_{\nu}$.
According to the above definition, the transforms
$\left\{F_{\nu}\right\}_I$ and $\left\{G_{\nu}\right\}_I$ already
contain an integration, therefore some terms in Eq. (8) present
twofold integration over 2D vectors ${\bf q}_{\lambda}$ and ${\bf
q}_{\nu}$. Really the latter, with the help of formula
$(2\pi)^{-2}\int\!\int d{\bf q}_1d{\bf
q}_2U(q_1,q_2)(q_{1+}q_{2-})^me^{\pm {i({\bf q}_1\times {\bf
q}_2)_z}}\!\equiv\!
\int_0^{\infty}\!\!\int_0^{\infty}dq_1dq_2(q_1q_2)^mU(q_1,q_2)
J_{\pm m}(q_1q_2)$ ($J_m$ is the Bessel function, $U$ is an
arbitrary function), is reduced to integration over absolute
values ${q}_{\lambda}$ and ${ q}_{\nu}$. Finally the numerical
summation over $n_{\nu}$ is performed.

In so doing, a simplifying circumstance was found: all of the
twofold-integration terms cancel each other in the final
combination (20). (This feature is not a general one but only
inherent in our specific case.\cite{foot}) All the rest terms
result in the following expression: \vspace{-2.mm}
$$
{}\!{}\!{}\!{}
\begin{array}{l}
\Delta E_{\mbox{\scriptsize
   SF}}=-\sum\limits_{n=2}^{\infty}S_n\frac{1-2^{1-n}}{n(n^2-1)},
\quad\mbox{where}\qquad{}\qquad{}\qquad{}\\{}\quad{}\qquad{}\qquad{}
S_n\!=\!\frac{2}{n!}\!\int_0^{\infty}dqq^{2n\!+\!3}
   V^2(q)e^{-q^2}.
\end{array}\eqno (21) \vspace{-2.mm}
$$
For the ideally 2D system we have $S_n\!\equiv\!1$, and the
summation may be easily performed, yielding $\Delta
E_{\mbox{\scriptsize SF}}=(\ln{2}-1)/2=-0.1534...$ (in units of
2Ry).

{\bf 5}. So, the shift is negative and the exchange interaction
lowers thereby the CSFE energy relative to the singlet MP mode.
The sign of the shift presents an expectable result. Indeed, the
second-order correction to the energy of a low-lying excitation
should be {\it presumably} negative due to the same reasons which
determine the {\it inevitably} negative sign of the correction to
the ground state energy. Another remarkable feature of the found
shift is its independence of the magnetic field.

Due to the ${\bf q}\!=\!0$ condition the studied state is
optically active and should be observed in photo-luminescent and
inelastic light scattering (ILS) measurements. In the recent work
\onlinecite{kul04} the ILS was studied in a single $30\,$nm
AlGaAs/GaAs QW in the situations where ${\mbox{\footnotesize
${\cal V}$}}\!=\!2; 4$. The triplet and MP cyclotron excitations
are manifested as peaks in the ILS spectra. The measurements were
performed in magnetic fields varied in a wide range, but with the
filling factor kept constant. The central triplet line is shifted
downward from the cyclotron energy by $0.35\,$meV independently of
the $B$ magnitude. Thus, a qualitative agreement with our
calculation is obvious.

Quantitative comparison should be done with taking into account of
finite thickness of a two-dimensional  electron gas. The
calculation in Fig. 1 incorporates the effect of the finite width
of the 2D layer. This is carried out by writing the Coulomb vertex
as $V(q)\!=\!F(qw)/q$, where the form factor $F(qw)$ is
parametrized by an effective thickness $w$. The latter
characterizes the spread of the electron wavefunction in the
perpendicular direction. If the variational envelope function is
chosen in the form $|\psi(z)|^2\!\sim\! \exp{(-z^2/2w^2)}$, then
$F(qw)\!=\!e^{w^2q^2}\mbox{erfc}(wq)$ (see Ref.
\onlinecite{co97}). Exactly this form factor is employed in the
calculation based on Eq. (21). Taking into account the value of
Ry$\!=\!5.67\,$meV in GaAs, we find from Fig. 1 that the agreement
with the experiment is obtained at $w\approx 0.5l_B$. This is
quite reasonable value for the $30\,$nm GaAs quantum structure.

 As a concluding remark we notice that the triplet
cyclotron excitation in spin-unpolarized electron system seems to
have been observed earlier, \cite{pi} although in this paper
experimental observations were related to the roton minimum and a
different experimental dependence of energy shift on magnetic
field was detected.

We acknowledge support by the Russian Fund of Basic Research. S.
D. thanks for hospitality the Max Planck Institute for Physics of
Complex Systems (Dresden), where an essential part of this work
has been done.

$\,$

$\,$

$\,$

$\,$

$\,$

$\,$

\begin{figure}[h]
\begin{center} \vspace{-8.mm}
\includegraphics*[width=0.7\textwidth]{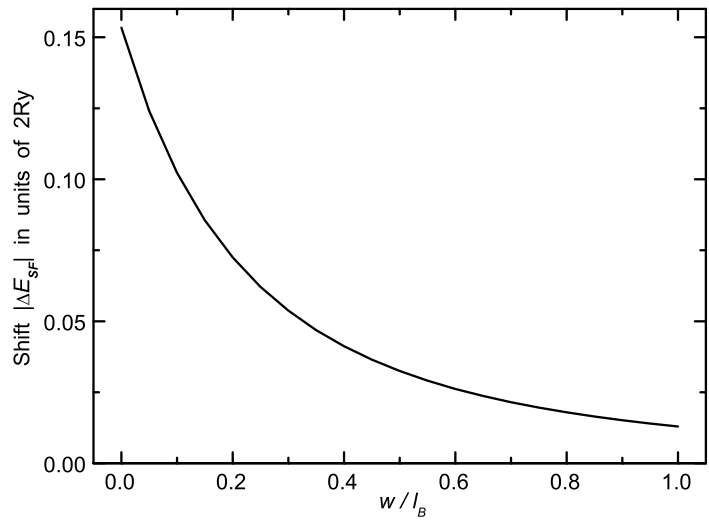}
\end{center}
\vspace{-2.6mm}
 \caption{The CSFE exchange shift is calculated from the
formula of Eq. (21) with the modified Coulomb interaction
$V(q)\!=\!q^{-1}e^{q^2w^2}\mbox{erfc}(qw)$; the shift value
absolute at $w\!=\!0$ is $(1-\ln{2})/2$.}
\end{figure}

%\end{thebibliography}


\begin{references}
\bibitem{foot}
{E.g. these terms are available in the second order corections to
the skyrmion-antiskyrmion gap [6] and to the energy of the hole at
the ${\mbox{\footnotesize ${\cal  V}$}}\!=\!1$ filling. We remark
that the hole energy correction was calculated in Ref. [6] without
taking into account the twofold-integration contribution. The
latter constitutes $-0.00349$ (in the strict 2D limit) and has to
be added to the result $-\pi^2/24\!+\!\frac{3}{4}(\ln{2})^2$
presented there. The total correction is thereby $-0.0544$ in
units of 2Ry (cf. the numerical result in Ref. [14]).}
\end{references}
\end{document}